# On the mechanism of high-temperature superconductivity in the iron based pnictides

K.P.SINHA

INSA Honorary Scientist, Department of Physics
Indian Institute of Science, Bangalore 560 012, India
(e-mail: kpsinha@gmail.com)

*Abstract:*

*The recent discovery of superconductivity at moderately high temperature (26 K to 55 K) in doped iron-based pnictides (Ln $O_{1-x}$ $F_x$ FeAs, where Ln = La, Ce, Sm, Pr, Nd, etc.), having layered-structure-like cuprates, has triggered renewed challenge towards understanding the pairing mechanism.*

*After reviewing the current findings on these systems, a theoretical model of a combined mechanism is suggested in which the phonon-mediated and distortion-field-mediated pairing processes give the right order of superconducting critical temperature $T_c$. The distortion-field modes arise from Jahn-Teller or pseudo Jahn-Teller effects due to degenerate or near-degenerate iron $3d_{xz}$ and $3d_{yz}$ orbitals.*

## Introductory Background

The discovery of iron-based pnictides, which can become superconducting at high temperature (26K to 55K range) on doping with fluorine e.g. Ln O FeAs → Ln $O_{1-x}F_x$FeAs (where Ln = La, Ce, Pr, Nd, Sm) has led to intense experimental and theoretical researches [1-14]. Like the cuprates, the structure of the new superconductors consists of layers. The layers of LnO alternate with layers of FeAs; the structure belongs to the tetragonal space group P4/nmn. It is believed that LaO bond is ionic and FeAs bond is largely covalent and the chemical formula is expressible as $(La^{3+}O^{2-})^+$ $(Fe^{2+}As^{3-})^-$ [2]. The degeneracy of 5(3d) orbitals of $Fe^{2+}$ is split by the crystal field. Early theoretical calculations gave the magnetic moment of iron as 2.00 $\mu_B$/Fe, but experiment at low T gives 0.35 $\mu_B$, and it may vary from 0.25 $\mu_B$ to 0.9 $\mu_B$ [15] Some workers claim that the observed magnetic moment may be explained by negative U by proposing exchange-induced effects for $Fe^{2+}$ $U_{eff}$ = U – J, where U is onsite Coulomb repulsion and J is atomic orbital intra-exchange energy (Hund parameter) [16]. Alternatively, the presence of magnetic frustration may give rise to reduced moment [13].

Neutron scattering results show that the parent compound LaOFeAs undergoes an abrupt structural distortion below 155 K from tetragonal to monoclinic [8]. At ~ 137 K it develops long-range antiferromagnetic order with small iron moment. Doping with fluorine suppresses both the magnetic order and structural distortion and superconductivity results. Owing to the $S_4$ symmetry of FeAs tetrahedra, the Fe 3d orbitals split into three non-degenerate ($3d_{x^2-y^2}$, $3d_{z^2}$, $3d_{xy}$) orbitals in both paramagnetic and antiferromagnetic states [17]. The splitting among the 3d orbitals is moderately small.

The observation that the appearance of superconductivity in the doped systems occurs with the disappearance of static antiferromagnetic order and lattice distortion rules out the mechanism based on spin density wave (SDW) and spin fluctuation exchange. The mechanism involving electron-phonon interaction has been found to be too weak to give the observed high $T_c$ [Boeri et al [18]]; however, Eschrig has presented a different picture [19]. Another, important experimental result is the enhancement of $T_c$ on applying pressure or replacing La by a smaller lanthanide ion such as Nd, Sm, etc., which will lead to shrinking of the lattice [2, 8].

The covalency is appreciable and one expects covalent tetrahedral bonding between hybridized Fe orbitals and As (4p) orbitals. The superconducting state is believed to have $d_{xy}$ orbital symmetry [14].

## Theoretical Formulation

The degeneracy or near degeneracy of $3d_{xz}$ and $3d_{yz}$ may be the source of lattice distortion via Jahn-Teller or pseudo Jahn-Teller effects [20-23]. The systems under study do exhibit structural instabilities due to tetragonal to orthorhombic distortion. This situation can be described by a two-level configurational distortion at each distorted tetrahedron at which the Fe ion exists.

A two-level system can be described by a pseudo spin (S = ½) formalism [24] because for every two-level system all 2x2 matrices and spin-half Pauli matrices along with 2x2 unit matrix form a complete set in the space of hermitian 2x2 matrices.

The Hamiltonian of a system of conduction electrons (in a narrow band) interacting with phonons and a distortion field (described by the two-level system) can be written as

$$H = H_e + H_c + H_{ph} + H_{e-ph} + H_{df} + H_{e-df} \quad , \tag{1}$$

where
$$H_e = \sum_{\mathbf{k},\sigma} E_k C_{\mathbf{k},\sigma}^+ C_{\mathbf{k},\sigma} \quad , \tag{2}$$

($C_{\mathbf{k}\sigma}^+$, $C_{\mathbf{k}\sigma}$) being the electron (creation, annihilation) operators for the state $|\underline{\mathbf{k}},\sigma\rangle$, $\underline{\mathbf{k}}$ is the wave vector, and $\sigma$ the spin index. $E_\mathbf{k}$ is the single-particle energy.

$H_c$ is the screened Coulomb interaction between conduction electrons; $H_{ph}$ is the standard phonon Hamiltonian; and $H_{e-ph}$ is the electron-phonon interaction. $H_{df}$ is the Hamiltonian of the distortion field that, in the pseudospin formalism, can be expressed as

$$H_{df} = -E_{as} \sum_m S_m^z - \tfrac{1}{2} \sum_{l,m} J_{lm} S_l^x S_m^x \quad , \tag{3}$$

where l and m are site indices. $E_{as} = E_a - E_s > 0$, being the energy difference between the two levels at site m, and $J_{lm}$ denotes the interaction energy between distortion at sites l and m.

$$H_{e-df} = \sum_{\underline{\mathbf{k}}-\underline{\mathbf{k}}'=\mathbf{q}} V_d(\underline{\mathbf{k}},\underline{\mathbf{k}}') C_{\mathbf{k}',\sigma}^+ C_{\mathbf{k}\sigma} S_q^x \quad , \tag{4}$$

which expresses the scattering of a conduction electron leading to a change in state of the distortion field.

The pairing coupling constant $\lambda_d$ involving the distortion field (df) mode can be calculated by using the method described by Vujicic et al. [25]. This yields

$$\lambda_d = N(o) <I_d^2> E_{as} <S^z> / \omega_d^2 , \tag{5}$$

where $\omega_d$ is the average frequency of the distortion-field mode. The phonon-mediated coupling constant is given by

$$\lambda_{ph} = N(o) <I_{ph}^2> <1/M\omega_{ph}^2> . \tag{6}$$

In equations (5) and (6), $<I_d^2>$ and $<I_{ph}^2>$ are the matrix elements of the corresponding interactions, $N(o)$ is the density of electronic states at the Fermi level, $\omega_{ph}$ is the phonon frequency, and M is the mass involved. We have also taken into account the role of phonons in the pairing process, though some workers have shown that the phonon contribution is too small [11, 18]. The vertex function here is a sum of phononic and df modes [25, 26].

The expression for the critical temperature for the combined mechanism turns out to be (after the standard calculation) [26]

$$T_c = 1.13 \, (\omega_{ph})^{r_p} (\omega_d)^{r_d} \exp(-1/(\lambda_{ph} + \lambda_d)) ; \tag{7}$$

where 
$$r_p = \lambda_{ph}/(\lambda_{ph} + \lambda_d), \quad r_d = \lambda_d/(\lambda_{ph} + \lambda_d) . \tag{8}$$

The superconducting gap at $T = 0$ is given by

$$\Delta(o) = \lambda_{ph} \int_0^{\omega_{ph}} \frac{d\varepsilon \, \Delta_{ph}(o)}{[\varepsilon^2 + \Delta_{ph}(o)^2]^{1/2}} + \lambda_d \int_0^{\omega_d} \frac{d\varepsilon \, \Delta_d(o)}{[\varepsilon^2 + \Delta_d(o)^2]^{1/2}} , \tag{9}$$

where $\Delta_{ph}(o)$ and $\Delta_d(o)$ denote the gaps for the pure-phonon and distortion-mode parts. This gives

$$\Delta(o) = \lambda_{ph}(o) \ln[\omega_{ph} + (\omega_{ph}^2 + (\Delta_{ph}(o))^2)^{1/2}/\Delta_{ph}(o)]^{\lambda_{ph}}$$

$$+ \lambda_d(o) \ln[\omega_d + (\omega_d^2 + (\Delta_d(o))^2)^{1/2}/\Delta_d(o)]^{\lambda_d} \tag{10}$$

The ratio obtained from (10) and (7) gives a value of $2\Delta(o)/T_c$ that is different from the BCS value of 3.52.

Numerical Estimates

To get an idea of the superconducting transition temperature $T_c$ we have computed it by using these values of the parameters involved, $\lambda_{ph} = 0.2$, $\lambda_d = 0.8$, $\omega_{ph} = (315.7)$ K. The value of $\lambda_{ph}$ assumed is almost the same as that evaluated by some workers [18]. The value of $\lambda_d$ (= 0.8) used here may change somewhat for different rare-earth ions or on application of pressure.

The value of $\omega_d$ will depend on the separation of the two levels in question. As the unit cell contracts on applying pressure or by replacing La with Sm, Ce, Pd, Nd ions, which have smaller ionic radii, $\omega_d$ will increase. The computed critical temperature values (for increasing values of $\omega_d$) turn out to be

| $\omega_d$ | 40 | 50 | 75 | 100 |
|---|---|---|---|---|
| $T_c$ | 25 | 29 | 42 | 52 |

## Concluding Remarks

The foregoing shows that, as in cuprates, a combined mechanism mediated by phonons and electron-induced-distortion-field mode can account for the high $T_c$ of the ferropnictides. In the case of cuprates, besides phonons the second mode involved lochons (electron pairs as local charged bosons) [26, 27].

## Acknowledgements

The author should like to thank Dr. A. Meulenberg for helpful discussions. This work is supported in part by Science for Humanity Trust, Bangalore, India and Indian National Science Academy.